\documentclass[epj,twocolumn]{webofc}
\usepackage[varg]{txfonts}   

\newcommand{\subtxt}[1]{\mbox{\scriptsize #1}}

\woctitle{19th Joint Workshop on Electron Cyclotron Emission and Electron Cyclotron Resonance Heating}
\begin{document}
\title{Perturbing microwave beams by plasma density fluctuations}
%
%

\author{\firstname{Alf} \lastname{Köhn}\inst{1}\fnsep\thanks{\email{alf.koehn@ipp.mpg.de}} \and
        \firstname{Eberhard} \lastname{Holzhauer}\inst{2} \and
        \firstname{Jarrod} \lastname{Leddy}\inst{3} \and
        \firstname{Matthew B.} \lastname{Thomas}\inst{3} \and
        \firstname{Roddy G. L.} \lastname{Vann}\inst{3}
}

\institute{Max Planck Institute for Plasma Physics, Garching, Germany
\and
           Institute of Interfacial Process Engineering and Plasma Technology, University of Stuttgart, Stuttgart, Germany
\and
           York Plasma Institute, Department of Physics, University of York, York, U.K.
          }

\abstract{%
The propagation of microwaves across a turbulent plasma density layer is investigated with full-wave simulations. To properly represent a fusion edge-plasma, drift-wave turbulence is considered based on the Hasegawa-Wakatani model. Scattering and broadening of a microwave beam whose amplitude distribution is of Gaussian shape is studied in detail as a function of certain turbulence properties. Parameters leading to the strongest deterioration of the microwave beam are identified and implications for existing experiments are given.
}
\maketitle
\section{Introduction}\label{s:intro}
Electromagnetic waves in the microwave regime are widely used for heating and diagnostic purposes in present fusion experiments based on magnetic confinement. In both cases, the microwaves must propagate across the plasma boundary, a region where substantial density fluctuation levels up to $100\,\%$ are known to occur~\cite{Zweben.2007}. These fluctuations can disturb a traversing microwave beam, basically changing its beam size and thus potentially spoiling heating efficiencies or leading to ambiguous diagnostics results.

Within this project, the perturbing effect of plasma density fluctuations on a propagating microwave beam is investigated by means of full-wave simulations. The advantage of a full-wave treatment as opposed to geometrical optics techniques is that no restricting assumptions about the size of the turbulent density structures or their amplitude need to be made. The disadvantage is the increased requirement for computational resources which is, however, not a limiting issue with the availability of powerful scientifically focussed computational facilities. 

Two full-wave codes are applied, IPF-FDMC~\cite{Koehn.2008} and EMIT-3D~\cite{Williams.2014}, which are both based on a cold plasma description. Parameter scans are performed in which the properties of the turbulence are varied. To be statistically relevant, the full-wave codes require an ensemble average over many density profiles, all having the same average turbulence properties. Such an ensemble of density profiles was created by a Hasegawa-Wakatani drift-wave turbulence model within the BOUT++ framework~\cite{Dudson.2009}. 

This work is the continuation of previous works in which the influence of single blob-like structures on a traversing microwave beam was investigated~\cite{Williams.2014,Koehn.2015}.

\section{The simulation set-up}
In this section, the full-wave codes and the computational grid are briefly described. This includes the technique how the fluctuating electron plasma density profiles are obtained.

\begin{figure}[b]
\centering
\includegraphics[width=.47\textwidth, clip]{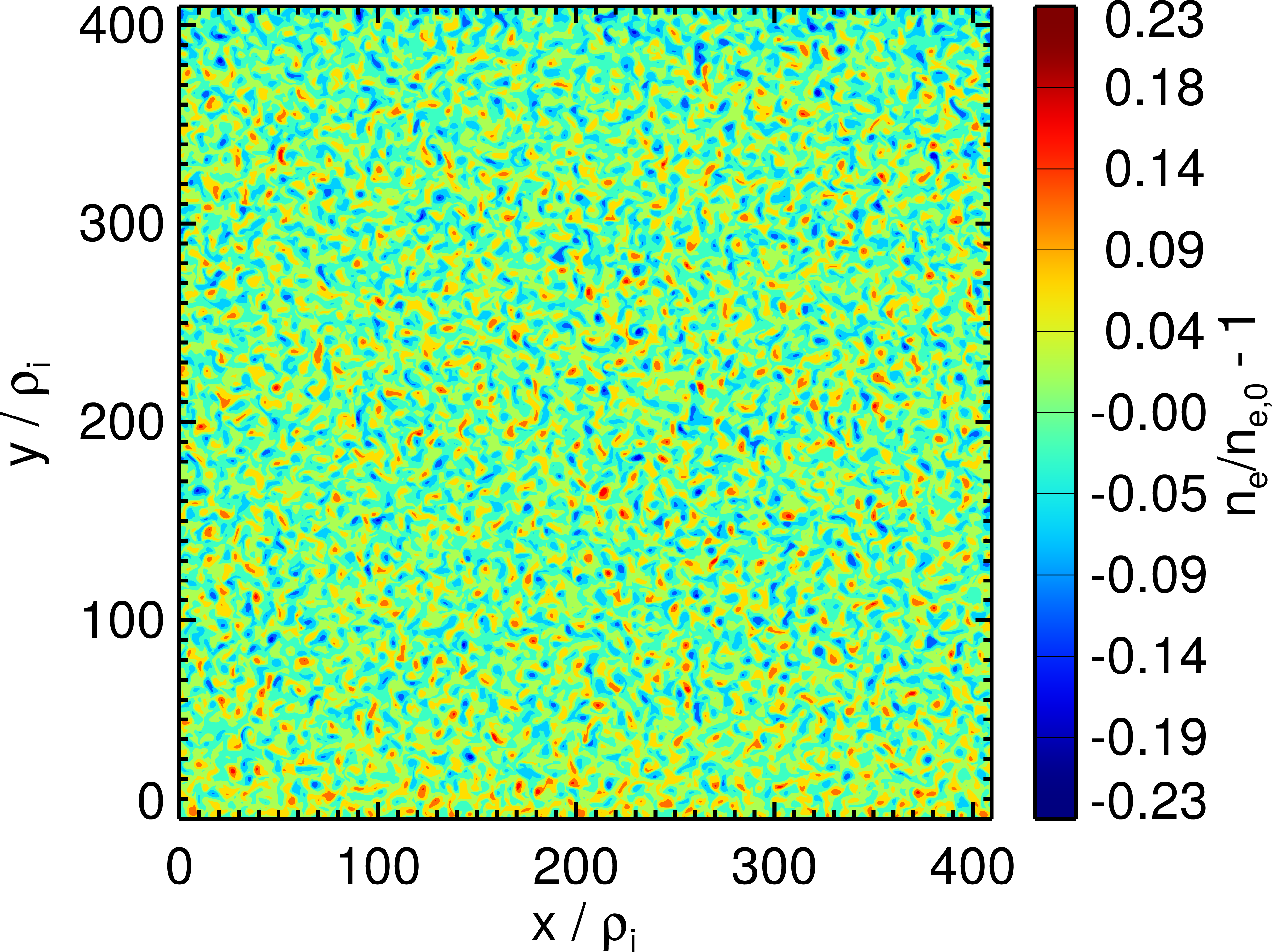}
\caption{Contour plot of the mean-free electron plasma density used as an input in the full-wave simulations. Shown is one time slice of the turbulence which evolves in time. The full set of turbulence is available at~\cite{Leddy.2016}.}
\label{f.BOUTpp}
\end{figure}

\subsection{The full-wave codes IPF-FDMC and EMIT-3D}
Both full-wave codes solve Maxwell's equations together with an equation for the current density in the plasma derived from the fluid equation of motion of the electrons:
\begin{eqnarray}
\frac{\partial}{\partial t} \mathbf{B} &=& - \nabla\times\mathbf{E} \\
\frac{\partial}{\partial t} \mathbf{E} &=& c_0^2\nabla\times\mathbf{E} - \mathbf{J}/\epsilon_0 \\
\frac{\partial}{\partial t} \mathbf{J} &=& \epsilon_0\omega_{pe}^2\mathbf{E} - \omega_{ce}\mathbf{J}\times\mathbf{\hat{B}}_0 - \nu_e\mathbf{J},
\end{eqnarray}
with $\omega_{pe}$ the electron plasma frequency, $\omega_{ce}$ the electron cyclotron frequency, $\mathbf{\hat{B}}_0$ the unit vector into the direction of the magnetic field, and $\nu_e$ the electron collision frequency. The equations are solved with the finite-difference time-domain (FDTD) method on a Cartesian grid (see e.g.\ Ref.~\cite{Taflove.2000} for a comprehensive overview of the FDTD method).

A Gaussian beam in O-mode polarization is injected into the grid using a so-called \emph{soft source}, i.e.\ the wave electric field is added to the grid at the position of the antenna. The amplitude distribution of the field added reads, neglecting the phase terms, $E(x)=\exp(-x^2/w_0^2)$, where $x$ is the coordinate across the antenna and $w_0$ the radius of the beam at the waist which is located in the antenna aperture. The antenna itself extends along the whole $x$-range of the grid to have a smooth field without the unwanted side lobes that would occur at truncated edges. If not mentioned otherwise, a value of $w_0=2\,\lambda_0$ is used with $\lambda_0$ the vacuum wavelength of the injected microwave.

\subsection{The plasma turbulence}
\begin{figure}[t]
\centering
\includegraphics[width=.47\textwidth, clip]{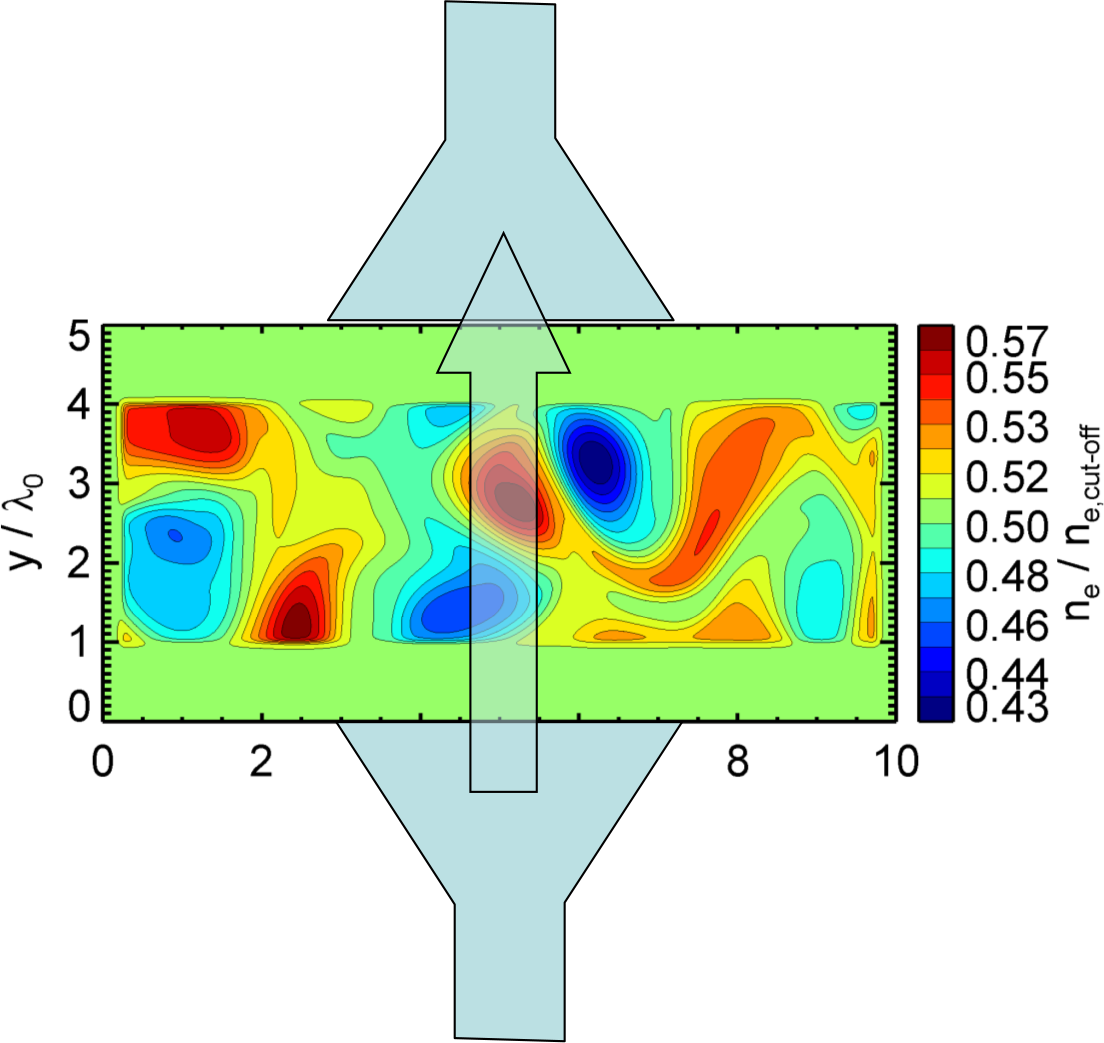}
\caption{Computational grid used in the full-wave simulations: a Gaussian beam in O-mode polarization is injected at the bottom and detected at the top after interacting with a layer of electron plasma density turbulence (the antenna structures are added for illustration purposes).}
\label{f.grid_antennas}
\end{figure}

The plasma density turbulence used in the full-wave simulations resembles the type of turbulence which is thought to be the dominant mechanism responsible for the anomalous transport observed in the edge of fusion plasmas, namely drift-wave turbulence~\cite{Horton.1990,Wootton.1990}. The Hasegawa-Wakatani description~\cite{Wakatani.1984} is used to model this turbulence within the BOUT++ framework~\cite{Dudson.2009}. Figure~\ref{f.BOUTpp} shows a snapshot of the density turbulence obtained from a BOUT++ run. The data is available for further simulations or benchmarks and can be freely accessed~\cite{Leddy.2016}.

Since a cold plasma description is used here, the microwave interacts only with the plasma density turbulence which is seen by the microwave as electron density fluctuations. In the time frame of the microwave, the fluctuations are frozen. This is due to the typical frequency scale of the fluctuations which lies in the kHz range (to be compared with the GHz range of the microwaves). In addition, the group velocity of the microwave is orders of magnitudes above the phase velocity of the density structures which can be approximated by the electron diamagnetic drift velocity and reaches values of $10^4\,\mathrm{m/s}$~\cite{Zweben.2007}. Therefore, the plasma density fluctuations are taken as fixed non-uniformities of the electron density in the full-wave simulations. 

\begin{figure}[t]
\centering
\includegraphics[width=.47\textwidth, clip]{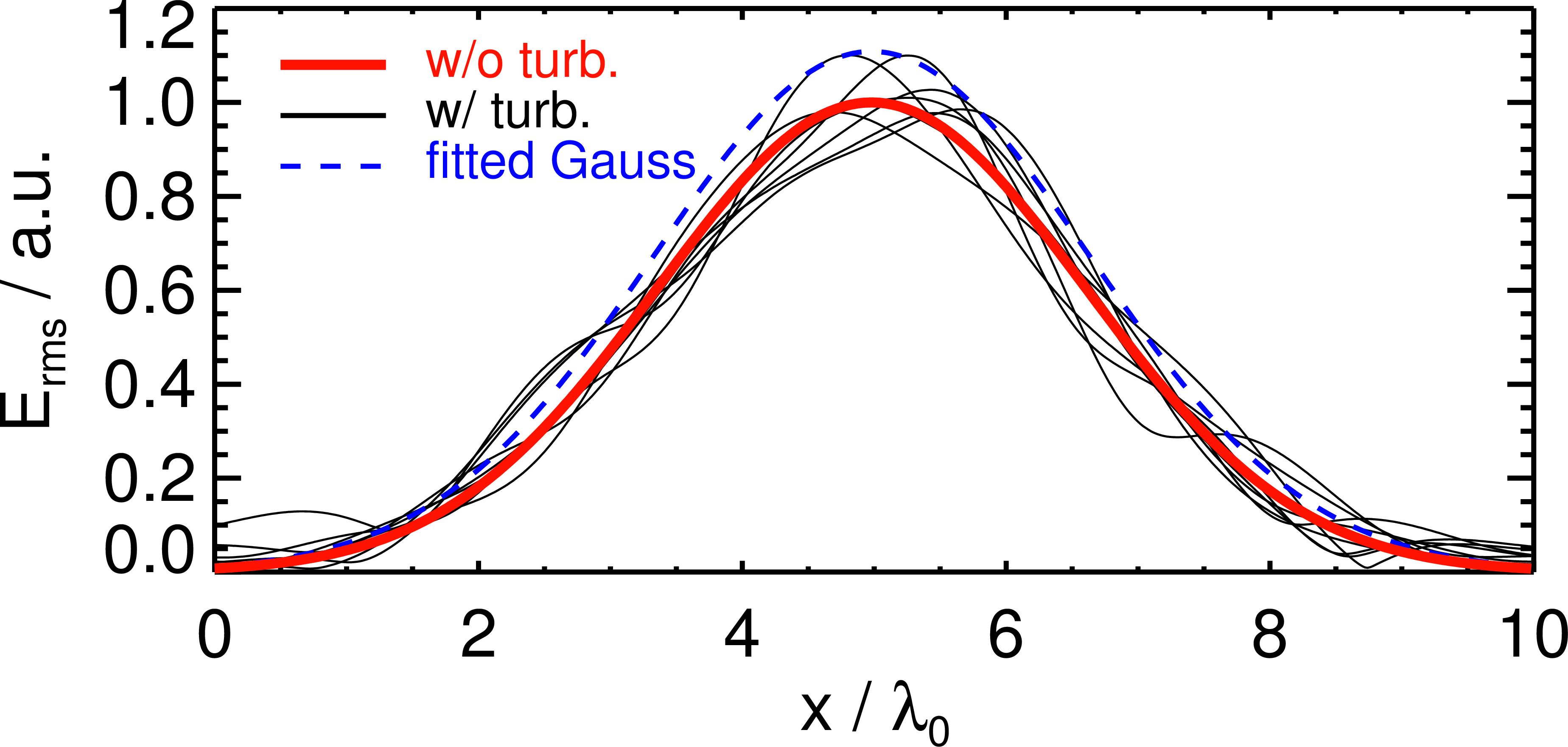}
\caption{Signal in the detector antenna plane for a few samples belonging to one ensemble for a set of turbulence parameters (black), for the homogeneous case (red), and a Gaussian fitted to all signals from this ensemble (dashed blue).}
\label{f.detAnt_signals}
\end{figure}

In order to accurately describe the experimental situation, it is not sufficient to perform a single full-wave run with one electron density profile. An ensemble of profiles is required to get statistically relevant results. Such an ensemble was generated with the BOUT++ code as described above~\cite{Leddy.2016}. From the spatially large snapshots of the BOUT++ runs, only a small area is needed for a single full-wave run (corresponding to one sample). Within one ensemble, the separate samples correspond each to an area cut from the full grid (the samples do not overlap).

\subsection{The computational grid}\label{s.setup_grid}
With IPF-FDMC, simulations are performed on a 2D grid which has a standard size of $10\times5$ vacuum wavelengths, as indicated in Fig.~\ref{f.grid_antennas}. An emitting antenna is located at the bottom at $y=0$ and a receiving antenna at the top at $y=5\,\lambda_0$. The grid is surrounded by non-radiating boundaries. As can be seen in Fig.~\ref{f.grid_antennas}, the grid has a homogeneous background electron density $n_{e,0}$ with a value of half of the O-mode cut-off density $n_{e,\subtxt{cut-off}}$. The fluctuations are actually located in a layer with a width of $3\,\lambda_0$, where a smooth transition from the homogeneous background to the turbulence layer is employed over a few grid points to avoid spurious reflections.

\begin{figure*}
\centering
\includegraphics[width=.97\textwidth, clip]{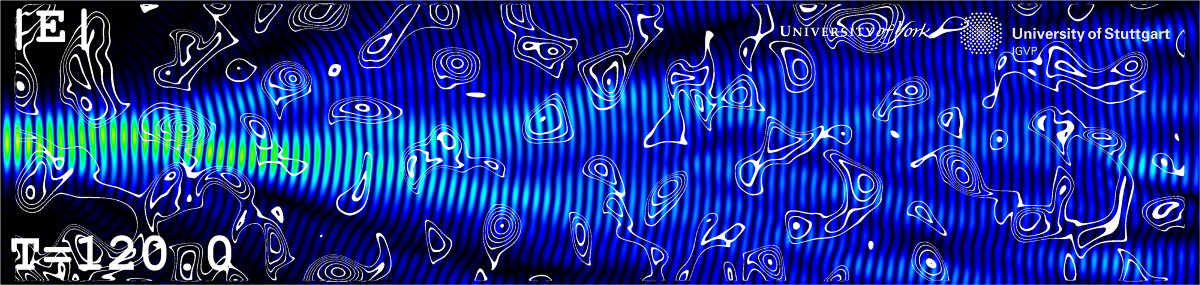}
\caption{Snapshot of the absolute value of the wave electric of an electromagnetic wave propagating across a plasma with electron density fluctuations, indicated by the white contour lines representing positive perturbations as compared to the background density. The snapshot is taken from a video published at~\cite{Koehn.2016}.}
\label{f.grid28}
\end{figure*}

The background magnetic field has a normalized value of $Y=\omega_{ce}/\omega_0=0.5$ and is directed perpendicular to the grid. Due to the drift-wave nature of the density structures, they are elongated along the magnetic field lines. This resembles the 2D grid used in IPF-FDMC, which assumes no variation in the third dimension. In EMIT-3D, however, full 3D runs are performed with the same grid size in the perpendicular direction (perpendicular to the background magnetic field). In the third dimension, the density structures are assumed to not vary and are thus simply extended along the magnetic field. This allows to compare the 2D simulations of IPF-FDMC with the 3D simulations of EMIT-3D and therefore elaborate if the scattering in the geometry used is basically 2D or 3D in nature. Due to the increased demand in computational resources when going from a 2D simulation to a 3D simulation, the comparison is only carried out for a few dedicated cases. In addition, a few 3D simulations are performed to investigate the effect of an oblique injection of the microwave beam onto the electron density filamentary structures.

\section{Data analysis}\label{s.data_analysis}
The full-wave simulations are based on a time-dependent scheme and start with the injection of the microwave beam at the bottom sending antenna. They are stopped when the microwave has propagated across the turbulence layer and a steady state solution is achieved. During the simulations, the wave electric field is continuously recorded at the receiving antenna, which spans across the whole simulation grid, see Fig.~\ref{f.grid_antennas}. More precisely, a time-averaged field is recorded, defined as follows: 
\begin{equation}
	\tilde{E}_{\subtxt{rms}}=\sum_t\frac{\sqrt{ \tilde{E}_x^2 + \tilde{E}_y^2 + \tilde{E}_z^2}} {\sqrt{T}},
	\label{e.Erms}
\end{equation}
with $t$ the time step in the full-wave simulations, $T$ the number of the wave periods passed since the start of the simulation and the superscript $\tilde{ }$ referring to the perturbations due to the electron density fluctuations. Figure~\ref{f.detAnt_signals} shows detector antenna signals according to Eq.~(\ref{e.Erms}) for a few samples from one ensemble of density profiles. The corresponding signal for the homogeneous case, i.e.\ without fluctuations is also shown, clearly illustrating the perturbing effect of the fluctuations. 

The scattering observed for each sample of density turbulence is quantified by summing up the squared deviations of the $\tilde{E}_{\subtxt{rms}}$ signal to the homogeneous case. Thus, a scattering parameter $\alpha$ is defined:
\begin{equation}
	\alpha = \frac{\sum_x\left( \tilde{E}_{\subtxt{rms}} - E_{\subtxt{rms}} \right)^2} {\sum_x E_{\subtxt{rms}}^2}.
\label{e.alpha}
\end{equation}
A separate value of $\alpha$ is obtained for each sample and ensemble averaging is then performed to get the scattering for one set of turbulence parameters.

A second analysis is applied which consists in determining the average detector antenna signal by calculating the arithmetic mean of the $\tilde{E}_{\subtxt{rms}}$ signals from all samples belonging to one ensemble. A Gaussian function is then fitted to the averaged signal, allowing to get the beam size of the averaged perturbed beam, $w_{\subtxt{turb}}$. Thus an average broadening of the microwave beam can be determined. This is important for the case of a localized current drive by injected microwaves with the idea of stabilizing neo-classical tearing modes (NTMs) which can lead to a sudden loss of the confinement in a tokamak~\cite{Poli.2015}.

\section{Simulation results}
Before starting to analyse the scattering described in the previous Section, an illustrative example is presented first. In contrast to the grid shown in Fig.~\ref{f.grid_antennas}, a much larger computational grid is used in order to emphasize the effect of electron density perturbations on a microwave beam. The background density corresponds again to half of the O-mode cut-off density. As shown in Fig.~\ref{f.grid28}, the scattering of the microwave can be clearly seen in this example, leading not only to a deviation of the original, straight beam path but also to a splitting into multiple beams. The simulation was obtained with the 2D full-wave code IPF-FDMC and can be accessed as a video, showing the microwave beam propagating across the grid, at Ref.~\cite{Koehn.2016}.

For the quantitative analysis of the scattering process, ensemble averaging is required as explained in Sec.~\ref{s.data_analysis}. To ensure that the ensemble is large enough, the average position of the maximum value of the $\tilde{E}_{\subtxt{rms}}$ signal in the detector antenna plane is compared with the position of the maximum for the homogeneous case. The difference should asymptotically approach zero (within error bars). It has been assured that this is the case for all ensembles considered in this paper. The resulting size of an ensemble can be as large as $N=25,000$. It varies when different average spatial sizes of the electron density structures are considered, where the spatial size is measured as the perpendicular correlation length $L_c$ which is varied from $L_c\approx0.06\,\lambda_0 \ldots 1.2\,\lambda_0$. With decreasing values of $L_c$, more density structures fit into the turbulence layer and the spatial average of one sample becomes better reducing the required ensemble size. 

\begin{figure}[t]
\centering
\includegraphics[width=.45\textwidth, clip]{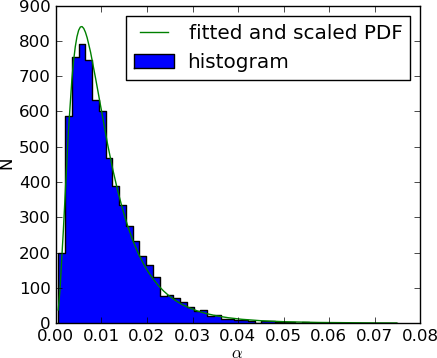}
\caption{Histogram and fitted PDF of the scattering parameter $\alpha$ for a size of the turbulent density structures of $L_c\approx0.5\,\lambda_0$. The histogram follows a log-normal distribution.}
\label{f.alpha_histogram}
\end{figure}

According to Eq.~(\ref{e.alpha}), $\alpha\ge0$ is always fulfilled. It is therefore not expected that $\alpha$ follows a normal distribution. Figure~\ref{f.alpha_histogram} shows the obtained distribution for a structure size of $L_c\approx0.5\,\lambda_0$: it actually follows a log-normal distribution. This applies to all values of $L_c$ and to statistically describe the scatter parameter $\alpha$, its median and the interquartile range will therefore be used.

\begin{figure}[t]
\centering
\includegraphics[width=.45\textwidth, clip]{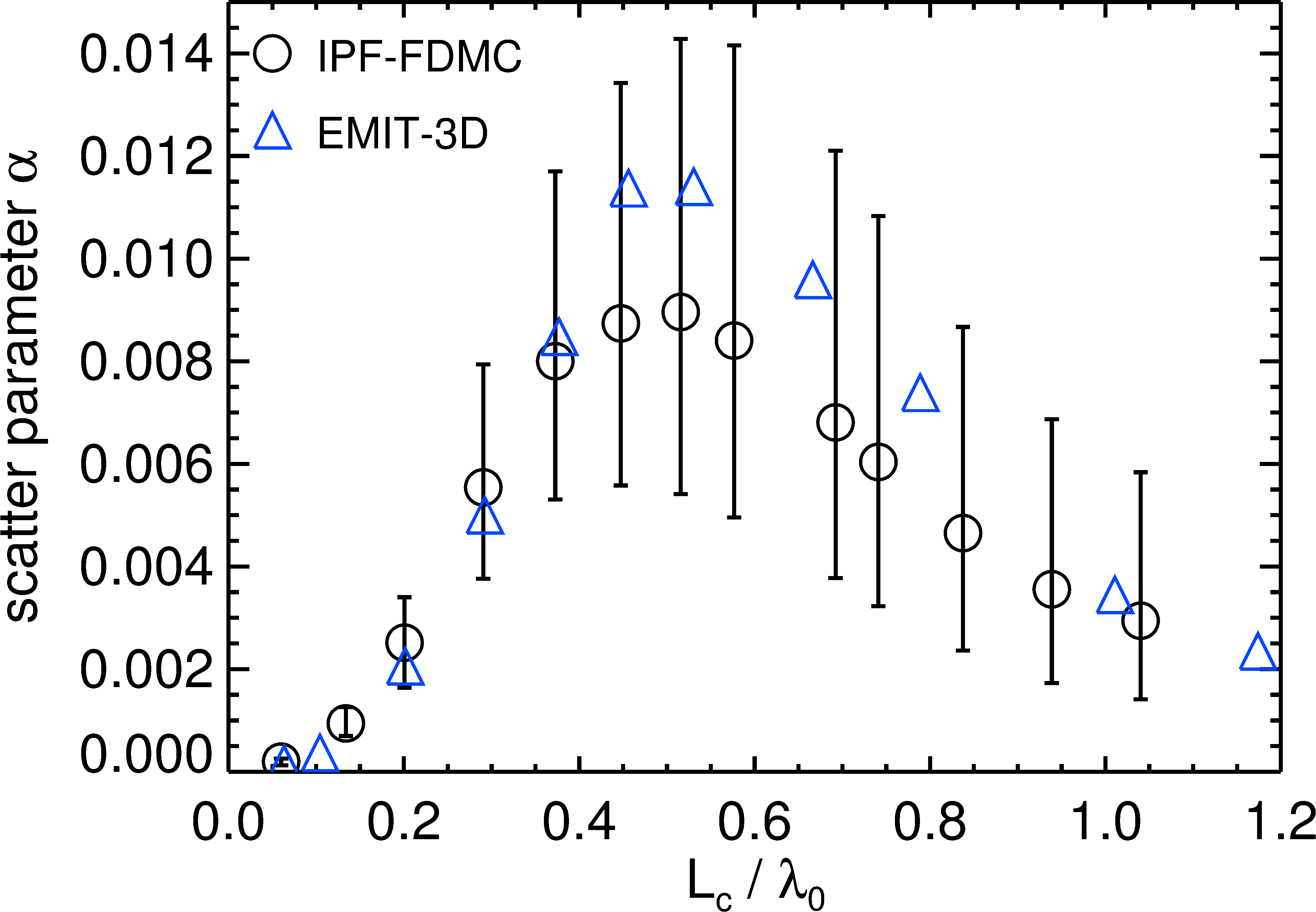}
\caption{Median of the scatter parameter $\alpha$, see Eq.~(\ref{e.alpha}), as a function of the electron density structure size for an average fluctuation strength of $\sigma\approx4\,\%$. The error bars correspond to the interquartile range.}
\label{f.Lc_scan_alpha3}
\end{figure}

Figure~\ref{f.Lc_scan_alpha3} shows the scattering as a function of the structure size for an average fluctuation strength of $\sigma\approx4\,\%$, which is measured as 
\begin{equation}
	\sigma=\sqrt{ \frac{1}{N_{x,y}} \sum_{x,y} \left( \tilde{n}_e(x,y)-n_0 \right)^2 },
	\label{e.flucLevel}
\end{equation}
with $N_{x,y}$ the number of grid points in the turbulence slice. Looking first at the results from IPF-FDMC, the scattering $\alpha$ exhibits a maximum at around $L_c\approx0.5\,\lambda_0$. Very large structures exceeding the width of the turbulence layer act effectively as a phase plate and the scattering is therefore expected to approach an asymptotic value as observed. If the density structures are too small they hardly have an effect on the microwave beam (obstacles with a size below $\lambda_0/10$ can generally be considered as not perturbing the microwave). The errorbars correspond to the interquartile range and obviously $\alpha$ has the largest spread where the maximum is located. 

The results from EMIT-3D are also shown in the plot and very good agreement is found. Note that the ensemble size is reduced by approximately two orders of magnitude as compared to IPF-FDMC due to the additional third dimension which leads to increased computational time. The agreement shows that the scattering in the geometry used, is basically a 2D process. 

\begin{figure}[t]
\centering
\includegraphics[width=.45\textwidth, clip]{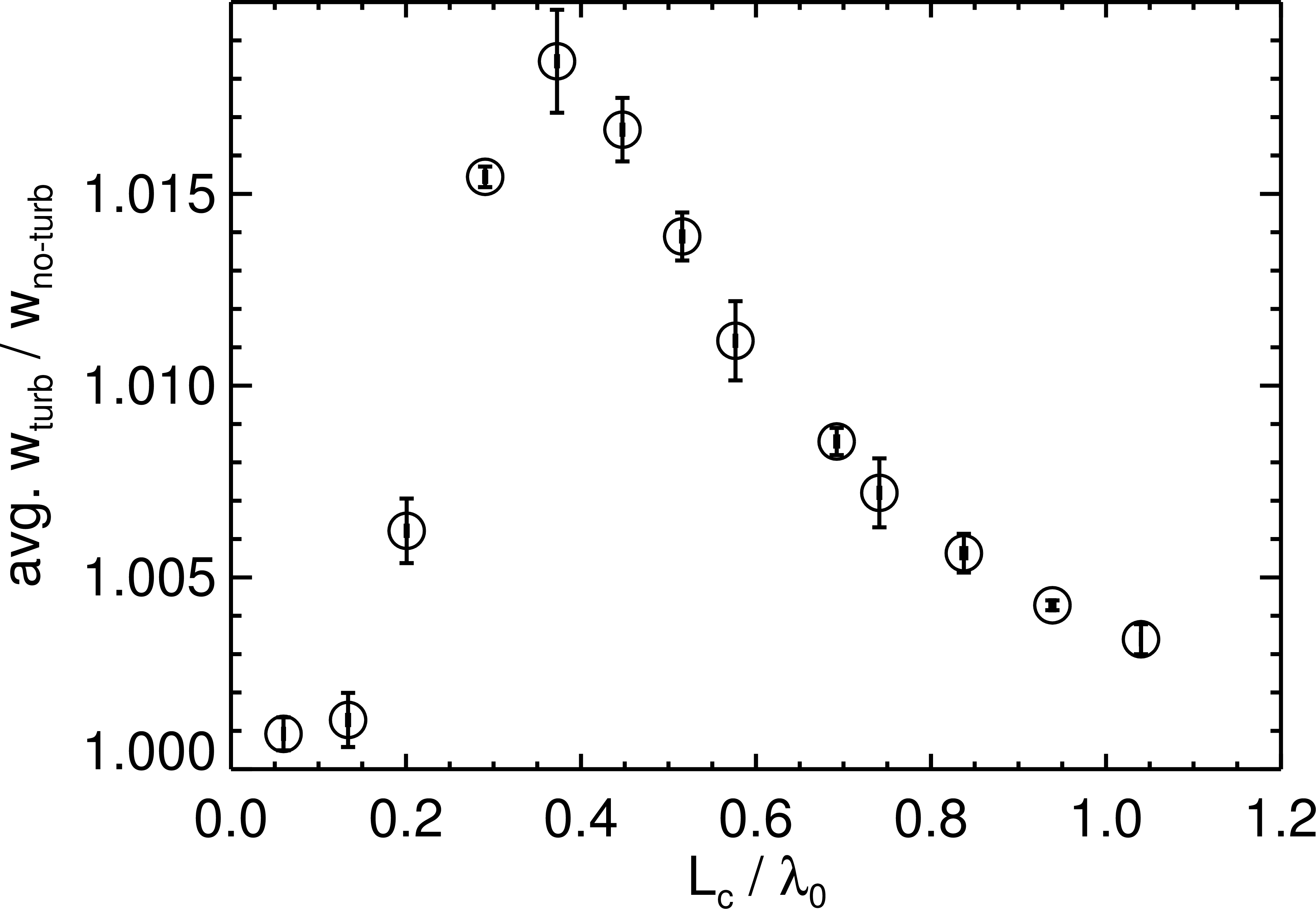}
\caption{Average beam broadening as a function of the density structure size for an average fluctuation strength of $\sigma\approx4\,\%$.}
\label{f.Lc_scan_alpha5}
\end{figure}

The average beam broadening obtained from the same simulations is shown in Figure~\ref{f.Lc_scan_alpha5} as a function of $L_c$. It follows a similar behaviour as $\alpha$, although the maximum is more pronounced. A broadening of approximately $2\,\%$ is found at maximum. 


Another important parameter influencing the scattering is the strength of the fluctuation amplitude. A number of parameter scans have been performed with the 2D full-wave code in which the fluctuation strength was varied in a range of $\sigma=2\ldots12\,\%$. The average structure size was $L_c\approx0.5\,\lambda_0$ in all cases, corresponding to the strongest scattering found in the previous parameter scans, see Fig.~\ref{f.Lc_scan_alpha3}. The width of the turbulence layer was kept constant. It is found that both $\alpha$ and the average beam broadening as a function of the fluctuation strength $\sigma$ follow a quadratic increase. 

In an additional scan, the width of the turbulence layer is varied in a range from $2\ldots7\,\lambda_0$. A linear influence of the width is found on both the scattering parameter $\alpha$ and the average beam broadening.

Further parameter scans performed in the same geometry are described in detail in Ref.~\cite{Koehn.2016b}.

\section{Application of simulation results}
Microwaves can be used to drive localized toroidal net currents in order to stabilize NTMs~\cite{Poli.2015}, as briefly mentioned earlier. The results obtained with the full-wave simulations are now applied to the case of the ASDEX Upgrade tokamak~\cite{Zohm.2007} for such a scenario. The microwave frequency is $140\,\mathrm{GHz}$, corresponding to a vacuum wavelength of $\lambda_0\approx2\,\mathrm{mm}$. The turbulence layer is located at the scrape-off layer with a density of approximately $10\,\%$ of the cut-off density of the injected microwave~\cite{Stober.2000} and an average size of the perturbing density structures of $L_c\approx4\,\lambda_0$~\cite{Fuchert.2014}. Although the average fluctuation level is with $\sigma\approx15\,\%$~\cite{Hennequin.2015} relatively large, only small perturbations of the microwave are expected due to the large density structure size with respect to the vacuum wavelength and the relatively low background plasma density.

\section{Summary}
Full-wave simulations of a Gaussian beam in O-mode polarization injected onto a layer of electron plasma density turbulence have been performed. It was shown that the scattering is basically 2D in nature when the microwave beam is injected perpendicular onto the plasma density structures which are elongated along the magnetic field lines, resembling drift-wave turbulence structures. The strongest deterioration of the microwave was found for an average density structure size corresponding to half of the vacuum wavelength of the injected microwave. A square dependence on the strength of the fluctuation and a linear dependence on the width of the turbulence layer was found in the parameter range used here. The simulation results were applied to one example, NTM stabilization in the ASDEX Upgrade tokamak, with the result that the deterioration of the microwave seems not to be important in this case.

\section{Acknowledgements}
One of the authors (A. K.) wants to thank Burkhard Plaum for providing him with a useful png-plotting library.

Part of the simulations were performed on the HELIOS supercomputer system at Computational Simulation Centre of International Fusion Energy Research Centre (IFERC-CSC), Aomori, Japan, under the Broader Approach collaboration between Euratom and Japan, implemented by Fusion for Energy and JAEA. 

One of the authors (M.B. T.) was funded by the EPSRC Centre for Doctoral Training in Science and Technology of Fusion Energy grant EP/L01663X.

%
%

\end{document}